# DesignCon 2012

A Zero Sum Signaling Method for High Speed, Dense Parallel Bus Communications

Chad M. Smutzer, Mayo Clinic
smutzer.chad@mayo.edu

Robert W. Techentin, Mayo Clinic

Michael J. Degerstrom, Mayo Clinic
degerstrom.michael@mayo.edu

Barry K. Gilbert, Mayo Clinic
gilbert.barry@mayo.edu, 507-284-4056

Erik S. Daniel, Mayo Clinic

## Abstract

Complex digital systems such as high performance computers (HPCs) make extensive use of high-speed electrical interconnects, in routing signals among processing elements, or between processing elements and memory. Despite increases in serializer/deserializer (SerDes) and memory interface speeds, there is demand for higher bandwidth busses in constrained physical spaces which still mitigate simultaneous switching noise (SSN). The concept of zero sum signaling utilizes coding across a data bus to allow the use of single-ended buffers while still mitigating SSN, thereby reducing the number of physical channels (e.g. circuit board traces) by nearly a factor of two when compared with traditional differential signaling. Through simulation and analysis of practical (non-ideal) data bus and power delivery network architectures, we demonstrate the feasibility of zero sum signaling and compare performance with that of traditional (single-ended and differential) methods.

## Author's Biographies

**Chad M. Smutzer** received a BSEE from the University of Iowa in Iowa City. He is currently a Senior Engineer at the Mayo Clinic Special Purpose Processor Development Group where he performs signal and power integrity analysis.

**Robert W. Techentin** received a B.S. degree in computer science from Rose-Hulman Institute of Technology in Terre Haute, IN. He is currently a Principal Engineer at the Mayo Clinic Special Purpose Processor Development Group. His research interests include software and hardware architecture, design, and development.

**Michael J. Degerstrom** received a BSEE from the University of Minnesota. Mike is currently a Senior Engineer at the Mayo Clinic Special Purpose Processor Development Group. His primary area of research and design has been in the specialty of signal and power integrity.

**Barry K. Gilbert** received a BSEE from Purdue University (West Lafayette, IN) and a Ph.D. in physiology and biophysics from the University of Minnesota (Minneapolis, MN). He is currently Director of the Special Purpose Processor Development Group, directing research efforts in high performance electronics and related areas.

**Erik S. Daniel** received a BA in physics and mathematics from Rice University (Houston, TX) in 1992 and a Ph.D. degree in solid state physics from the California Institute of Technology (Pasadena, CA) in 1997. He is currently Deputy Director of the Special Purpose Processor Development Group, directing research efforts in high performance electronics and related areas.

# Introduction

Many digital systems such as High Performance Computers (HPCs) make extensive use of both differential and single-ended channels and busses.  For example, high-speed differential serial channels are often used for processor-processor communications, while dense single-ended busses are typically used for processor-memory interfaces.  Differential signaling has many advantages (e.g. common-mode noise rejection, reduced simultaneous switching noise (SSN), etc.), but uses twice the number of interconnect traces as single-ended signaling. We propose that an intermediate solution might be optimum. Consider a set of buffers driving 2N interconnect traces, where at any given time, N of these traces are in a logic high state (sourcing current), and the other N of these traces are in a logic low state (sinking current), but without the restriction that a differential system would have (i.e., that the high and low state pairs must be adjacent). We call this "zero sum signaling" as it preserves one of the most important features of a differential buffer system – the constant supply current sourcing independent of the output states.  This constant current feature translates into greatly reduced switching current transients interacting with system inductance, and hence lower power supply voltage transients and associated data corruption. Using single-ended traces and a suitable coding scheme to create a balanced (or even nearly-balanced) number of logic high and low states across the bus at any point in time, it is possible to transmit more data down the 2N traces than would be possible with a differential signaling protocol, approaching the single-ended limit as the bus size (N) grows.

We begin this paper with a brief theoretical discussion of the zero sum signaling method.  We then detail the coding methodology and generate sets of balanced and nearly balanced code words.  Next, using time-domain simulations, we evaluate the feasibility of zero sum signaling as applied in a notional, HPC environment including realistic active and passive elements.  Within the simulation section, we explore a number of variations to the operating conditions and observe the impact to the high-speed link performance.  Performance is evaluated using nominal and more stressful bit pattern stimulus conditions across a parallel bus while monitoring metrics such as vertical eye opening at the end of the links.  The paper is completed with conclusive remarks and suggestions for additional effort.

# Zero Sum Signaling Basic Principles of Operation

Three different parallel bus data transmission methods are considered throughout this paper; single-ended (SE), differential (DIFF) and zero sum (ZS).  Given a fixed number of printed wiring board (PWB) traces, the number of data bits that can be transmitted using the three signaling schemes is explored in Table 1 below.  Note that the formula shown in the table for the number of zero sum bits comes from counting the number of codes available with 2N total bits in which N bits are ones and N bits are zeroes.  As N grows, the number of zero sum data bits approaches the single-ended data bit limit.

| Number of Traces (2N) | Single-ended Data Bits (2N) | Differential Data Bits (N) | Zero Sum Data Bits ($\log_2[(2N)!/(N!N!)]$) |
|---|---|---|---|
| 4 | 4 | 2 | 2.58 |
| 8 | 8 | 4 | 6.13 |
| 12 | 12 | 6 | 9.85 |
| 16 | 16 | 8 | 13.65 |
| 32 | 32 | 16 | 29.16 |
| 64 | 64 | 32 | 60.67 |

**Table 1: Comparison of Single-ended, Differential, and (optimal) Zero Sum Signaling Schemes.**

Obviously, for binary transmission, the non-integer ZS data bits in Table 1 are non-physical. Alternatively, we can effectively invert these formulae to compute the number of traces that would be required to carry a fixed number of bits using these signaling schemes. The results of these formulae are showing in Table 2 below. For the zero sum case, we restrict ourselves to an even number of traces (as otherwise an equal number of ones and zeros is not possible), computing the smallest even integer number of traces which can support the given number of data bits.

| Number of Data Bits | Single Ended Traces | Differential Traces | Zero Sum Traces |
|---|---|---|---|
| 8 | 8 | 16 | 12 |
| 12 | 12 | 24 | 16 |
| 16 | 16 | 32 | 20 |
| 20 | 20 | 40 | 24 |
| 24 | 24 | 48 | 28 |
| 32 | 32 | 64 | 36 |

**Table 2: Traces Required for Various Signaling Schemes In Order to Support a Fixed Number of Data Bits.**

A block diagram comparing notional single-ended, differential, and zero sum links for a 32 bit data bus is shown in Figure 1.

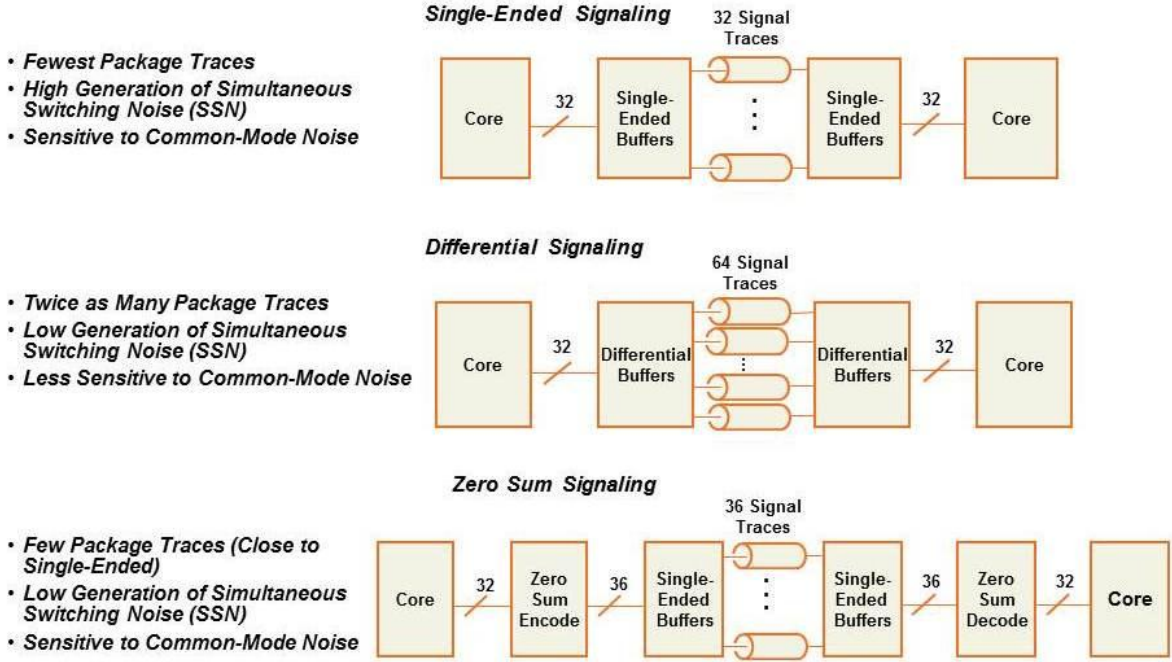

**Figure 1: Comparison of Traditional Signaling Schemes (Single-ended and Differential) and Zero Sum Signaling. (23489v3)**

Note that this concept can be extended to incorporate "nearly zero sum" encoding schemes. In this case, the number of zeroes and ones is not restricted to be equal, but may differ by some finite "disparity", d (e.g. d=±2, ±4, etc.). This situation would result in finite current switching and hence increased SSN, but with the benefit of an expanded set of code words, and hence more theoretically available bits for a given number of wires. Even though we are expanding the notion of zero sum signaling to encompass nearly zero sum codes, we will continue hereafter to use the term "zero sum signaling" to refer to the general concept in both cases, using the term "disparity" to reflect the allowed difference between the number of zeroes and ones across the zero sum bus at any given instant in time.

The formula in Table 1 for computation of the number of (encoded) bits which can be carried across a given number of traces can be extended to accommodate finite disparity as follows. The number of code words available across 2N bits which have exactly N-k ones and N+k zeroes (i.e. the code words with a disparity equal to 2k) can be computed as follows:

$$Codes(k) = \frac{(2*N)!}{[(N-k)!*(N+k)!]}$$

**Equation 1**

The same formula applies to the number of code words with N+k ones and N-k zeroes. One computes the number of effective bits available through the use of all code words by adding up the total number of code words with disparity, d, less than or equal to a given value and taking the log base two of this number. For example,

$$Bits(d=4) = Log_2[Codes(0) + 2*Codes(1) + 2*Codes(2)]$$

**Equation 2**

In Table 3 below, we apply these formulae to compute the number of data bits which can be supported across a range of assumed physical interconnect traces using zero sum signaling with a range of allowable disparities (abbreviated as "ZS ±d"). Note that in parenthesis, we represent the integer part of this computed number, as generally one is only interested in using a bus to transmit an integer number of bits per word. For comparison, the number of single-ended (SE) and differential (DIFF) data bits are shown as well.

| Number of Traces | SE Data Bits | DIFF Data Bits | ZS±0 Data Bits | ZS±2 Data Bits | ZS±4 Data Bits |
|---|---|---|---|---|---|
| 8 | 8 | 4 | 6.13 (6) | 7.51 (7) | 7.89 (7) |
| 12 | 12 | 6 | 9.85 (9) | 11.29 (11) | 11.77 (11) |
| 16 | 16 | 8 | 13.65 (13) | 15.13 (15) | 15.66 (15) |
| 20 | 20 | 10 | 17.50 (17) | 18.99 (18) | 19.56 (19) |
| 24 | 24 | 12 | 21.37 (21) | 22.88 (22) | 23.47 (23) |
| 32 | 32 | 16 | 29.16 (29) | 30.69 (30) | 31.32 (31) |

**Table 3: Data Bits Supported for Various Signaling Schemes Across a Fixed Number of Traces.**

Again, we can invert these formulae to compute the number of traces that would be required to carry a fixed number of bits using any of these signaling schemes. The results of these formulae are showing in Table 4 below. As before, we compute the smallest even integer number of traces which can support at least the given number of data bits. Note that allowing finite disparity does open up the code space such that additional bits can be carried, but that the restriction to even integer numbers of traces is such that the incremental code space allowed by ZS±4 does not allow a reduction in trace count relative to ZS±2 (at least for the cases considered in the table).

| Number of Data Bits | SE Traces | Diff Traces | ZS±0 Traces | ZS±2 Traces | ZS±4 Traces |
|---|---|---|---|---|---|
| 8 | 8 | 16 | 12 | 10 | 10 |
| 12 | 12 | 24 | 16 | 14 | 14 |
| 16 | 16 | 32 | 20 | 18 | 18 |
| 20 | 20 | 40 | 24 | 22 | 22 |
| 24 | 24 | 48 | 28 | 26 | 26 |
| 32 | 32 | 64 | 36 | 34 | 34 |

**Table 4: Traces Required for Various Signaling Schemes (Including Zero Sum with Finite Disparity) In Order to Support a Fixed Number of Data Bits.**

Before proceeding with the simulations, it was necessary to investigate the methods of generation of zero sum coded data, as will be described in the next section. The following sections will then describe the simulation details and conclusions.

# Zero Sum Coding

In order to implement the zero sum signaling concept, either in simulation or hardware, it is necessary to encode and decode arbitrary data words to and from a set of balanced code words, where the code words contain an equal number of zero and one bits. Substantial research has been done on balanced encoding and decoding [1], [2], [3], [4], [5] with a variety of applications, including data transmission and storage. Balanced coding for data transmission generally supports the goals of transition density and DC-free characteristics that are desirable for both electrical and optical links. Some coding techniques also address data integrity and error detection and correction.

Popular transmission encoding techniques such as 8b10b address the needs of a stream of bits transmitted over a single serial channel. The 8b10b code is not perfectly balanced, but ensures that the disparity in a stream of 20 bits is no more than 2, and that there are no more than five ones or zeroes in a row. Zero sum encoding, on the other hand, must address a parallel data "word," ensuring that there are an equal number of zero and one bits simultaneously transmitted over a data bus. Therefore, alternative encoding/decoding schemes and alternative hardware implementations are needed for zero sum encoding.

For the purposes of this paper, the main requirement was that a method of providing coded data for simulations could be developed. This task is relatively straightforward, as there were no particular limitations on the computing resources that could be employed for this purpose. Support of coding for simulation purposes is discussed in the next section. Of course, practical implementation of zero sum signaling would require an efficient hardware implementation of an efficient coding scheme. This is a much more difficult problem to solve, and was beyond the scope of the work performed for this paper.

**Coding Methods Used for Simulations**

The circuit simulations described in the upcoming analysis sections require substantial sets of random data for the "typical" cases. "Worst" case patterns are contrived in the dark hearts of engineers with knowledge of the weaknesses of the system. Random data for single links or uncorrelated single-ended links can be provided by the pseudo-random bit stream (PRBS) functions available in the simulation environments. Random streams of balanced code words, however, require specific efforts.

A simple encoding method could employ a lookup table of code words, where each data value represents an index into the lookup table. Random code words are generated simply by randomly selecting entries from the lookup table.

This is very simple to implement when encoding small data sets, but it becomes unwieldy as the number of data bits increases, and decoding requires searching the table. For the purposes of these simulations, however, the lookup table is both appropriate and efficient. For small sets of code words, the entire table is generated. For large sets of code words, the lookup table is populated with randomly chosen balanced (or nearly balanced) code words. Since the simulations require (at most) hundreds of code words, lookup tables on the order of 10,000 entries are more than adequate in size, and easily generated or stored.

Whether code word values for the lookup table are enumerated or randomly chosen, candidate code words are converted to binary, and the zero and one bits are counted. Balanced code words have an equal number of one and zero bits. As described above, "nearly balanced" code words, with bounded disparity, are identified by the difference, d, in the number of 0 and 1 bits. For example, for the case of $2*N = 4$ (wires), code words are 4 bits and there are $2^4 = 16$ of them. The code words for 4 bits and their disparities are enumerated in Table 5. (Note that the number of codes in each column can be derived from Pascal's Triangle.)

| Disparity (d) | -4 | -2 | 0 (Balanced) | +2 | +4 |
|---|---|---|---|---|---|
| Code words | 0 0 0 0 | 0 0 0 1 | 0 0 1 1 | 0 1 1 1 | 1 1 1 1 |
| | | 0 0 1 0 | 0 1 0 1 | 1 0 1 1 | |
| | | 0 1 0 0 | 0 1 1 0 | 1 1 0 1 | |
| | | 1 0 0 0 | 1 0 0 1 | 1 1 1 0 | |
| | | | 1 0 1 0 | | |
| | | | 1 1 0 0 | | |

**Table 5: Disparities for 4-bit Code Words.**

Matlab software code was developed to generate sets of code words with specific code sizes and disparities, and then to randomly generate code words to serve as input to the simulations. For example, consider the case of N=8, for which zero sum encoding requires 12 wires (as opposed to 16 for differential signaling or 8 for single-ended signaling). Running the Matlab software for 40 words of 12 bit codes produces the result shown in Figure 2. Each column in the result represents one balanced code word, with six zeroes and six ones. Each row represents the bit stream to be transmitted by the simulated output buffer.

```
Encoded Data for 40 words of 12 bits

Data Words: 2676 221 1733 2227 2649 1596 1581 918 2676 492 1484 3880 3366 238

3722 1692 1475 811 1896 3876 3117 907 2482 1649 567 1141 3102 2701 2606 2502

3366 1449 2676 485 3672 1148 1926 845 1310 634

Each Row Represents One Channel
0111101000000001100110111010001010000100
0001000100001110110001101010111000001011
1110011111011010001100011111110110111110
0100111001110111011011000011100100110111
```

```
110111011000000100000011111000001011001 1
100101101101110001111011110010111101000 1
111010001110010010100001010001001111010 1
011100010110011100001100001010101001000
000000010111100011110110000001110100111 0
101011111001001101110101100110001010110 1
001001100011101110111001011000110011101 0
100110001001101000011010001111101010000 0
```

**Figure 2: Example of Matlab-Generated 12-bit Balanced Code Words.**

# Simulation Environment and Associated HPC System Assumptions

In this section, we describe the detailed model used for evaluating the zero sum signaling concept against traditional differential and single-ended alternatives. For the purposes of this discussion, it is assumed that the reader has a fundamental understanding of SSN and its impact to signal integrity in a traditional HPC system, as well as the basic principles of differential vs. single-ended signaling. A non-exhaustive discussion of SSN and the basic modeling and simulation principles can be found in [10],[11],[12],[13].

The theory behind the zero sum signaling concept, as explained in previous sections, is relatively straightforward yet there are limited examples defined in practice [14], [15]. In brief, the concept proposes the use of coded words to achieve current cancellation in the transmitting circuit and hence a reduction in SSN by suppressing current transients (di/dt). Under ideal conditions, current cancellation can clearly reduce SSN simply based on the fundamental L*di/dt relationship. But a typical HPC application is far from ideal. Reflections, loss, packaging parasitics, power/ground pin ratio and power/ground pin distribution are all potentially realistic HPC system attributes that could decrease the effectiveness of a zero sum signaling implementation depending on how the buffer currents actually propagate.

The simulated system elements and some description of the simulation embodiment are depicted and described in Figure 3. For the purposes of these simulations, the models and assumptions which comprise the simulated environment apply equally well to either processor-to-processor or processor-to-memory links. The arbitrary design goal was to provide 32-bit-wide links between each pair of processing nodes. These could be provided using differential signaling (requiring 64 wires), single-ended signaling (requiring 32 wires), or zero sum signaling (requiring between 34 and 40 wires, depending on the level of allowed disparity, and depending on the number of wires that were grouped together into a single zero sum bus). The I/O buffers were grouped into sections ("slices") as shown in Figure 3, with each slice capable of supporting up to 40 single-ended or 20 differential buffers. This partitioning was adopted loosely from an Altera Stratix IV FPGA chip and package architecture, adopted primarily for PDN modeling purposes. The implications of this architecture will be further discussed below.

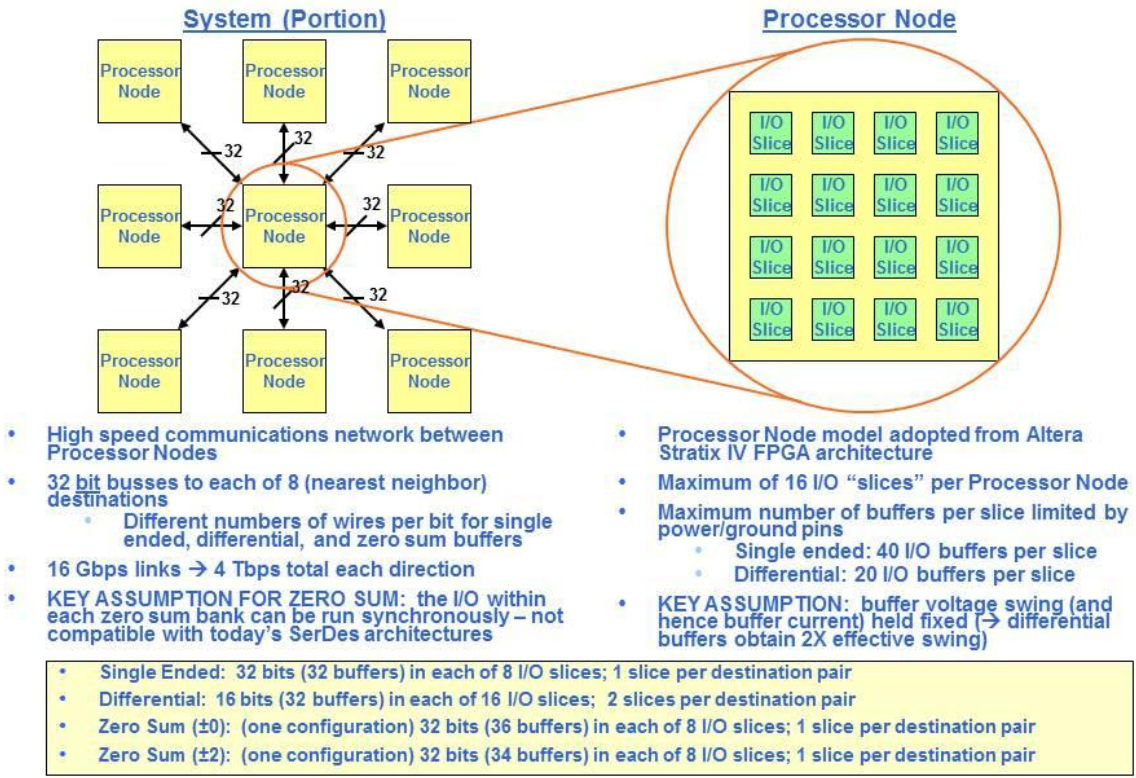

**Figure 3: Notional System Concept Used for Evaluating the Zero Sum Signaling Concept in the Context of Processor-to-Processor Interconnects. (41134v2)**

In this section, the theory is applied in a simulation environment to further understand the concepts and potential limitations of zero sum signaling in a more "real world" application. The simulation environment was created with the goal of comparing the performance of three signaling link architectures; single-ended (SE), zero sum (ZS) and differential (DIFF). The simulation environment for these architectures is described in detail in the following subsections along with simulation results and conclusions.

## **Link Architecture**

In an effort to emulate a more realistic application environment, the simulation construct shown in Figure 4 (representing the entire set of system components associated with one "slice" of the processor node, as shown in Figure 3 above) was created for each of three signaling architectures. For the purposes of these simulations, we assumed that the salient system features which would be relevant to zero sum signaling evaluation would be essentially independent from one slice to the next, so only a single slice was simulated. The complete simulated slice is depicted in Figure 4. This detailed figure describes many aspects of the high-speed link simulation architecture and it will be referred to often. The basic environment consists of three primary building blocks; the transmit buffer, the on-die power distribution network (PDN) and the passive printed wiring board (PWB) channel.  Each portion of the overall link structure, as outlined by Figure 4, will be described in the upcoming subsections.

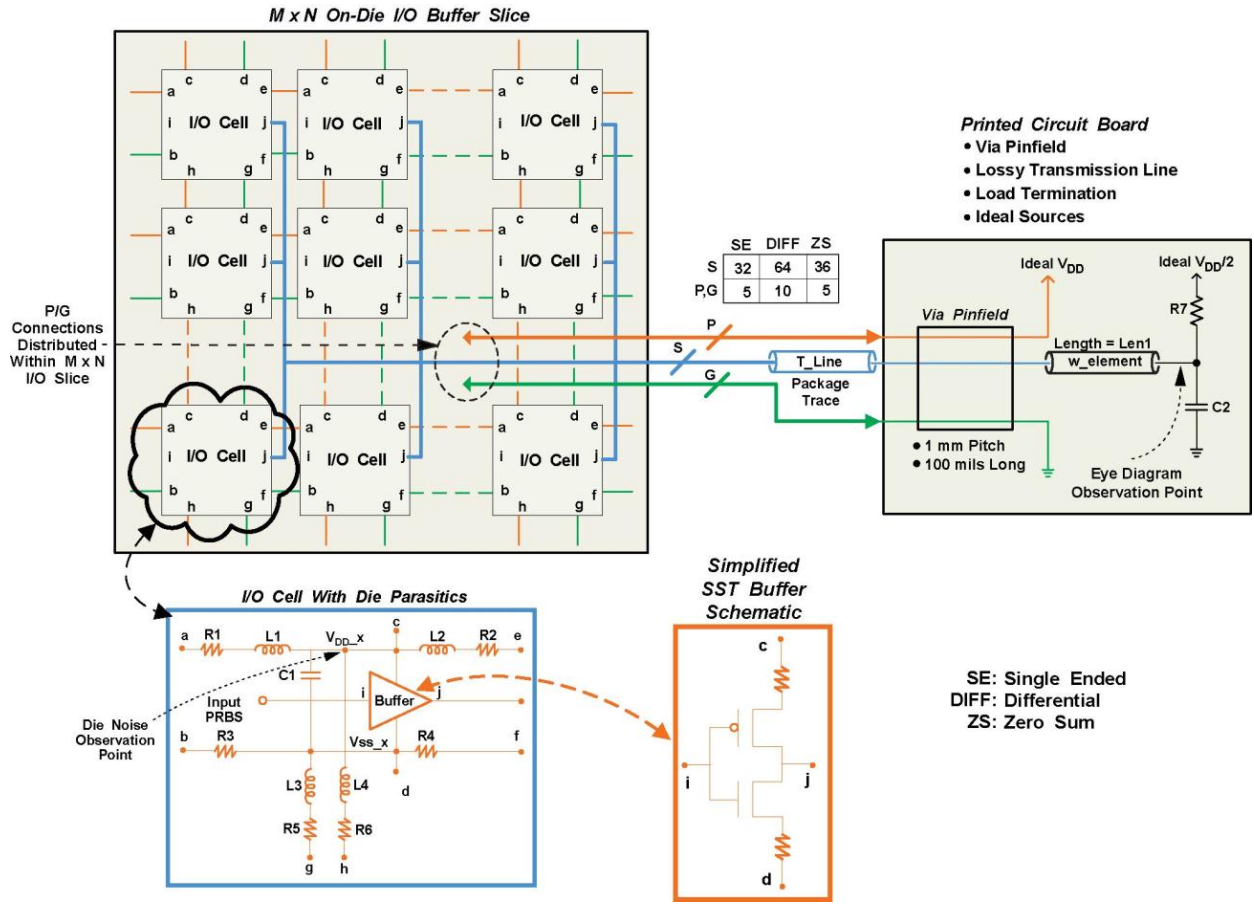

**Figure 4:  One "Slice" of Simulation Architecture Used For Evaluating Zero Sum, Differential and Single Ended Buffer Configurations and Impact to Simultaneous Switching Noise.  (41122v2)**

## **Link Architecture: Transmit Buffer**

The link begins with a transistor model for a push-pull style, source-series terminated (SST) output buffer, as is used in many serializer/deserializer (SerDes) implementations.  This transmitter has a 50 Ohm series output impedance by design and operates un-terminated from 0 to 1 VDC full swing and from approximately 0.25 to 0.75 VDC terminated with a matching 50 Ohm impedance.  The choice of this buffer was influenced primarily by prior experience [16], and because of the need for high-speed capability to stress the frequency-dependent channel and PDN.

In most differential-mode applications, a true current-steering differential buffer would be used to realize SSN immunity.  However, to promote consistency across the three architectures for this comparative study, two copies of the single-ended SST buffer were used to approximate one differential transmitter. Using two of these buffers does not ensure traditional SSN immunity that can be achieved with current-steering and may have resulted in overly pessimistic simulation results for the differential case. Therefore, it is noted that the differential conditions described herein are only quasi-differential in that not all traditional benefits obtained from practical differential signaling have been modeled. However, for the purposes of these simulation studies, we believe that this shortcoming is not of major significance. As will be shown in the simulation results below, the simulation models are able to clearly reflect differences in current switching and resultant SSN between single-ended and differential signaling, and hence we believe the zero sum results presented here are valid.

## **Link Architecture: On-Die Passive Power Distribution**

On-die I/O power distribution was modeled as an RLC network, shown in the lower left of Figure 4, intending to emulate the parasitics resulting from metal and dielectric materials within an FPGA or ASIC design. The passive RLC parameters used for this simulation exercise were estimated using a series of simplifying assumptions regarding the on-chip geometry and construction. For example, it was assumed that the metal strapping for on-chip power and ground could be modeled using traditional transmission line approximations such that:

$$Zo = \frac{\sqrt{\varepsilon_r}}{C * c}$$

**Equation 3**

$$Zo = \sqrt{\frac{L}{C}}$$

**Equation 4**

where Zo represents the characteristic impedance in Ohms, $\varepsilon_r$ is the dielectric constant, c is the speed of light in a vacuum, and L and C are the inductance and capacitance per unit length respectively for the power/ground network. Further assuming a power/ground characteristic impedance of 100 Ohms (a roughly chosen value, representing unintentionally coupled, physically separated power and ground traces), dielectric constant of 4 (representative of typical CMOS back-end-of-line dielectrics) and a buffer-to-buffer pitch of 190 um (typical die bump pitch), the lumped-model values for L and C were computed using Equation 3 and Equation 4 and documented in Table 6. On-die resistance was approximated using metal geometry and material assumptions from an IBM 45 nm chip technology. (Reference designators in Table 6 correlate to values in Figure 4.)

| Parameter | Value |
|-----------|-------|
| L1- L4 | 62 pH |
| C1 | 0.012 pF |
| R1-R6 | 0.35 Ω |

**Table 6: On-chip PDN RLC Parameter Definition.**

Some of these geometric assumptions may seem a bit pessimistic as compared to modern chip designs. For example, current ASIC designers are cognizant of SSN issues and therefore often add a great deal of on-chip capacitance to mitigate voltage rail collapse. However, in this study it was necessary to create an underperforming PDN such that signal degradation would be more obvious and relative comparisons between signaling styles could be made. Therefore, a somewhat poor PDN network was needed, and hence large amounts of on-chip capacitance were not included. Indeed, we treated several of these parameters as adjustable means for improving or degrading PDN performance.

The notional system example presented in Figure 3 with 32-bit wide busses between components was implemented as the baseline for comparison between SE, DIFF and ZS signaling. With the standard I/O cell defined in the lower left of Figure 4, a slice of cells were grouped together as shown by the generic MxN matrix in the upper left of Figure 4 for each of the three architectural test cases; 32 SE cells, 64 DIFF cells (32 differential pairs) and 36 ZS cells. The physical arrangement of the I/O cells within each slice, as well as the distribution of P/G pins, mirrored the PWB via pinfield distribution and is discussed in greater detail in the next section.

## **Link Architecture: Printed Wiring Board**

Perhaps the most influential component within the simulation architecture is the PWB via pinfield, due to its significant contribution of inductance of the signaling loop. Pin pitch, length and diameter, as well as the ratio between and location of signal and reference pins, all factor into the total loop inductance that directly contributes to SSN. The via pinfield I/O and P/G distribution modeled for each of the three test cases is shown in Figure 5. Although discussed in the context of the PWB via pinfields, the same organizational layout was repeated for the MxN matrix of on-chip I/O and P/G cells. The arrangements, especially the P/G-to-signal ratio, are loosely based on a reference design from Altera for a Stratix IV FPGA. The complex arrangement found in the Altera design was simplified using a more symmetric approach such that the pinfield could be easily scaled between the three architectures while maintaining some geometric consistency. Pin pitch was assumed to be 1 mm and the via lengths were modeled as 100 mils, each of which are fairly typical of modern ASIC or FPGA package and PWB implementations. Every signal, power and ground pin was routed through the via pinfield to the I/O slice with quantities of each described by the table in Figure 4 and further illustrated by Figure 5.

It is noted that these pinouts were not necessarily optimized for the least amount of SSN possible; especially with respect to the differential implementation. Alternative pinouts may result in better performance; however this approach was taken for sake of consistency between all three architectures so that comparative observations could be made.

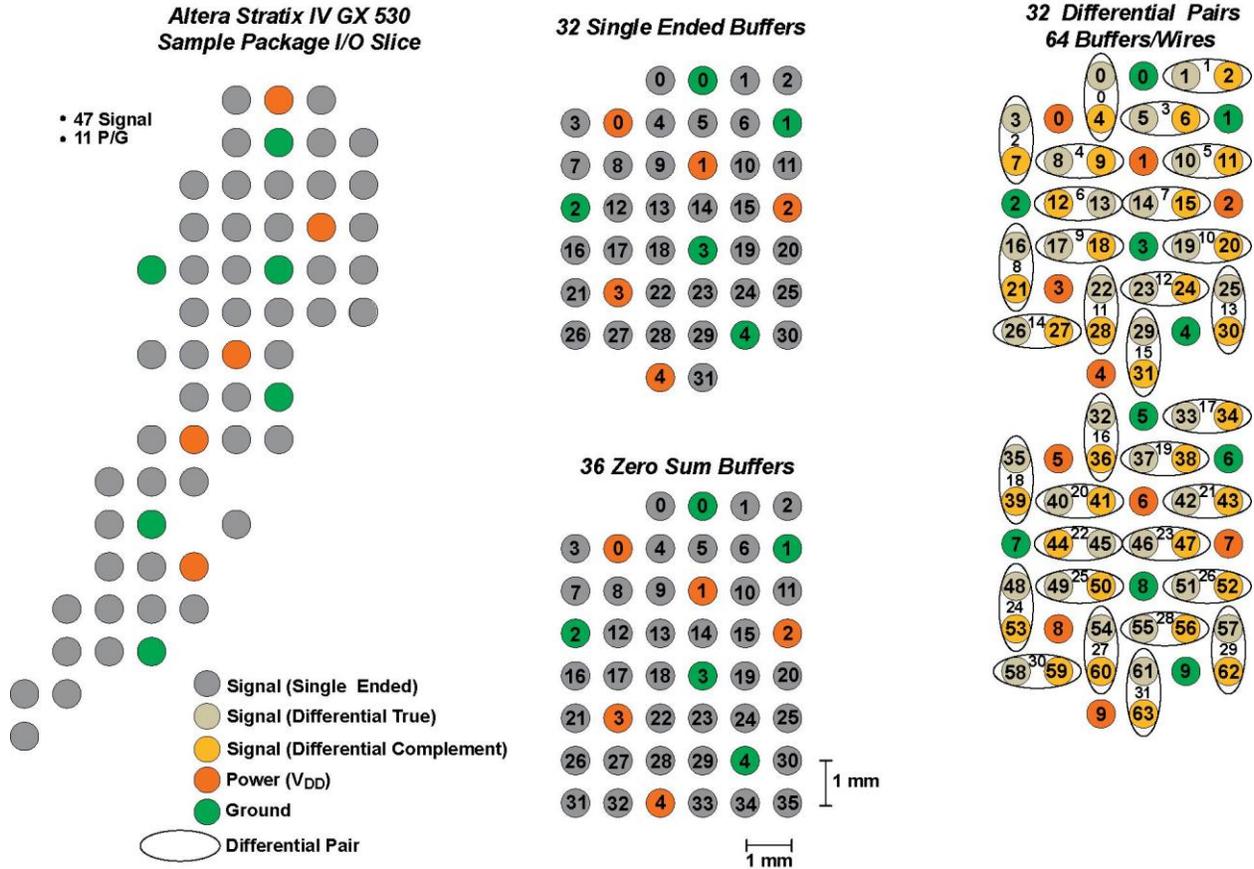

**Figure 5: Via Pinfield Assignments Describing Power, Ground and Signal Distribution for Three Architectural Configurations. (41291)**

The simulation link structure was completed with a lossy transmission line w-element model of length Len1, a termination resistor R7 and a small load capacitance C2, as shown on the upper right of Figure 4. To emulate variability in these PWB parameters, the length of the transmission line and value of R7 and C2 were randomly assigned from a specified range of typical values (Table 7) for each I/O channel. Transmission line length was constrained such that true-to-complement skew within a differential pair (relevant to the DIFF case) was kept below +/- 20 mils.

| Parameter | Range |
|---|---|
| Tline Length (Len1) | 3.98 – 4.02 in |
| R7 | 49 – 51 Ω |
| C2 | 0.3 – 0.5 pF |

**Table 7: PWB Link Parameter Range Values.**

Line-line crosstalk within the transmission line media was deliberately not modeled, as separate, uncoupled w-element models were used for each I/O cell. This was done in an effort to focus on signal degradation due primarily to SSN by removing other independent sources of noise. Using this approach, however, common-mode noise that would otherwise be generally rejected within tightly coupled pairs may instead have minimally contributed to eye closure in the differential measurements.

**Link Architecture: Simulation Stimulus and Patterns**

SSN is also pattern-dependent because the pattern of switching buffers across a slice influences local di/dt. A "worst case" switching pattern attempting to maximize di/dt was implemented by forcing all buffers within the slice to change to the same high or low state simultaneously. Nominal, or "typical", patterns for SE and DIFF were created by staggering the $2^7$-1 pseudo-random bit stream (PRBS) seed such that the signal edges were still aligned, but the distribution of 1's and 0's across the bus at any instance was more random in nature.

The concept of zero sum signaling ensures an equal number of zeroes and ones across the bus, thus global di/dt is zero, but localized stress on the PDN was achieved in the "worst case" condition by grouping the same-state transitions in each half of the total 36 buffer distribution. Finally, the "typical" ZS patterns were generated using randomly chosen valid ZS code words, as described in the coding methodology sections above. Each of the "worst case" and "typical" patterns for all three notional architectures is further illustrated in Figure 6. In the simulation environment, an ideal source defined by these pattern characteristics was used to stimulate the I/O cell at node 'i' as shown in Figure 4. Note that in all cases, clocking of the various patterns across the myriad of buffers was assumed to be synchronous (i.e., all buffers change state at node "i" simultaneously).

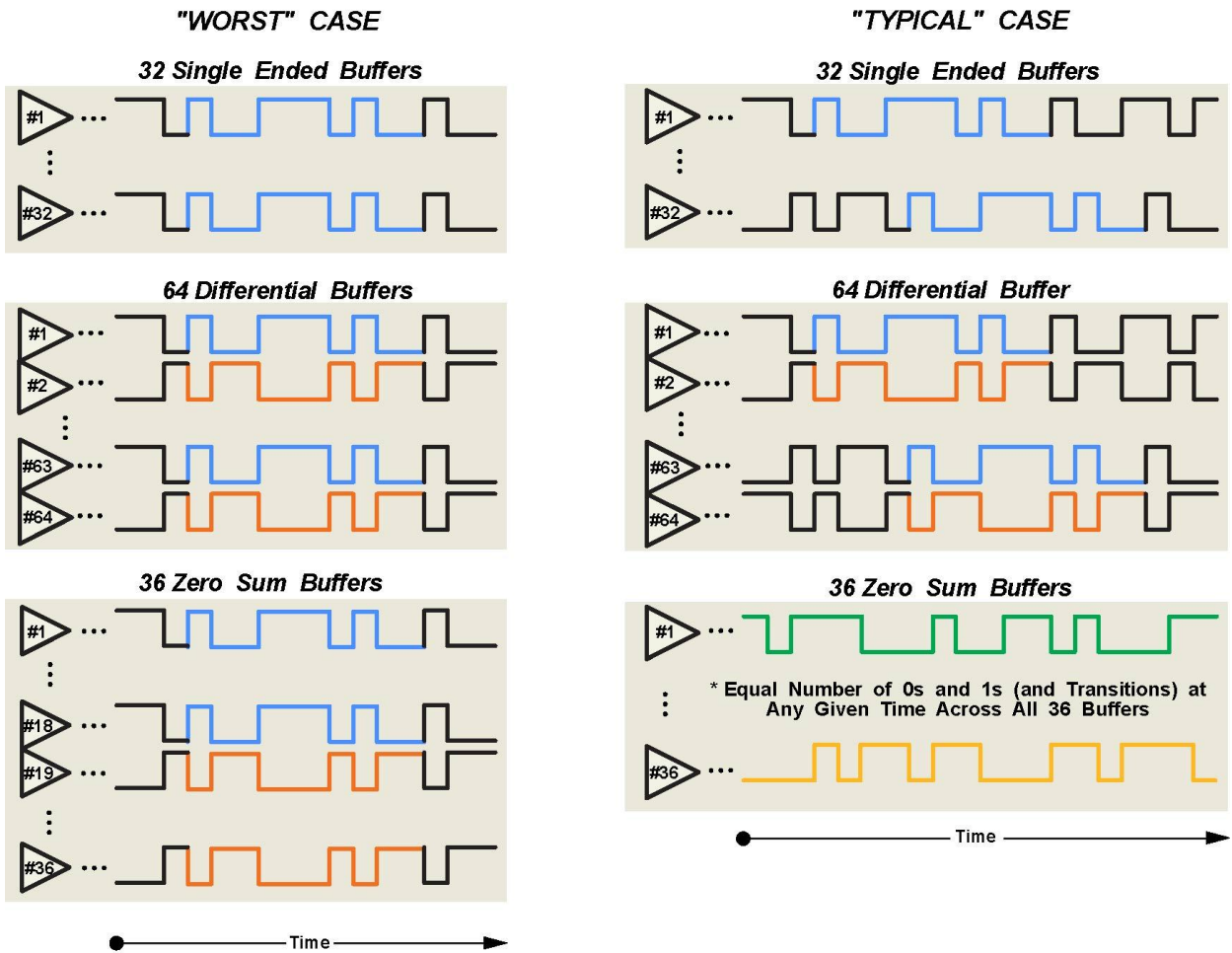

Figure 6:  Pattern Stimulus Description and Nomenclature Used for SE, DIFF, and ZS SSN Simulations. (41289)

As was discussed above, the arbitrarily selected target nodal bandwidth was approximately 4 Tbps for each processing element, as illustrated in Figure 3.  This bandwidth is hypothetically achieved with 8 lanes of 32 bits with each link (containing as many traces as needed in each case to represent a 32 bit word) operating at 16 Gbps.  Therefore, in the time-domain simulations, the nominal SST buffer input stimulus data rate was set to 16 Gbps for each transmitter (except for certain simulations, as noted below).  This speed was consistent with the particular SST buffers chosen for these simulations, as they were (though a past effort) designed to support 20+ Gbps speeds [16].

## Simulation Results

In this section, we summarize the results of numerous simulation studies of zero sum vs. differential vs. single-ended signaling, implemented using the model environment described above.  Synposys HSPICE version C.2009.03.SP1 for Linux was used for all time-domain simulations.  Matlab scripts developed by Mayo SPPDG were used to capture and measure pertinent eye diagram statistics.

## Nominal Simulations: Eye Diagrams and Voltage Ripple

The primary metric by which the three architectures were evaluated was vertical eye diagram opening as simulated/sampled at the receiver input. Therefore, other classical eye diagram characteristics, such as jitter, typically evaluated with an eye mask were not thoroughly considered in these simulation results. In typical high density, single-ended, parallel bus applications, SSN can be large and the effects are visible in the signal eye diagram. To illustrate these SSN and other effects, sample eye diagram simulation results are presented in Figure 7. Buffer location 14 for the SE and ZS matrix and buffers 14/15 for DIFF were arbitrarily chosen for this example.

As mentioned above, the rail-rail voltage swing for the SST buffer is nominally 0 to 1 VDC. However, in practice the buffer is terminated to $V_{DD}/2$ or 0.5 VDC. With this termination approach, the maximum ideal full voltage swing for each single-ended link measured at the terminating resistor is 500 mV. The DIFF configuration results are reported as a differential measurement where the complement signal was subtracted from the true signal within a pair. In this case, the nominal full swing eye is twice the single-ended nominal or 1 VDC. Note the 2x vertical scale in the differential eye openings in Figure 7.

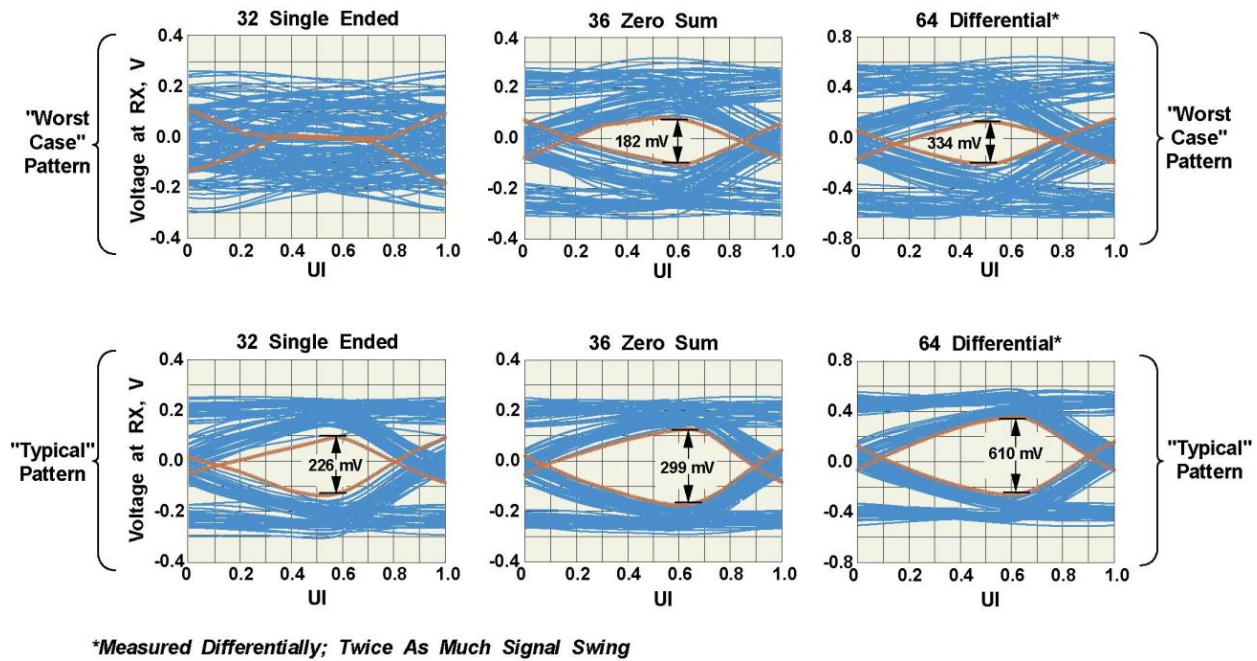

**Figure 7: Sample Eye Diagrams For Each of Six Configurations In 16 Gbps Simulations. (41294)**

The vertical axis is labeled "Voltage at RX" and is meant to represent the eye voltage present at the receiver input (Figure 4 shows exactly where these eye diagrams were measured in the simulation). Also noteworthy is the zero common-mode voltage for the ZS and SE measurements which is simply an artifact of the eye diagram plotting tools used for this study (which help facilitate comparison between the three signaling methods). In reality, the SE and ZS buffer voltage swings have a common-mode voltage roughly 500 mV.

An obvious observation from the plots in Figure 7 is the relative eye closure of the "worst case" results (top row) as compared to the "typical". This clearly demonstrates the pattern dependent behavior and resulting impact of same-state, simultaneous signal transitions within a grouping of high-speed I/O. However, a single sampling within the slice is not a complete picture of the performance across the entire bus. The eye opening for each SE, ZS buffer or DIFF pair within their respective I/O slice is presented in Figure 8. In this plot, each small horizontal line represents the vertical eye measurement of a location within the slice. Additionally, the minimum, average and maximum eye opening for each configuration is also labeled.

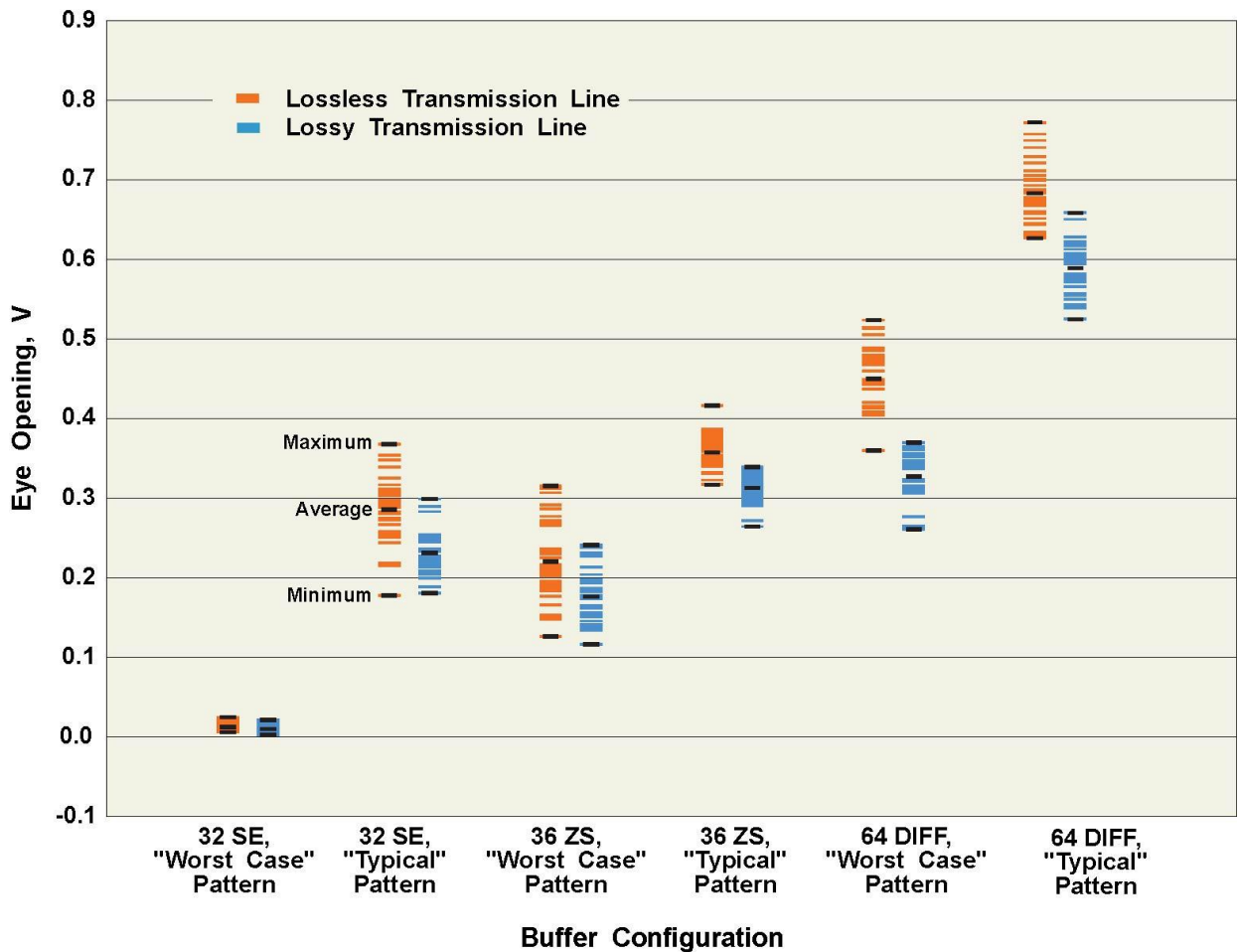

**Figure 8: Vertical Eye Opening Summary for Three Buffer Configurations and Two Switching Patterns In a 16 Gbps Simulation. (41292)**

As expected, the eye diagrams for the single-ended, "worst case" arrangement are completely closed. The SSN is too great and the voltage rail collapse prevents the buffers from operating full swing. The differential eyes are considerably more open and could likely be resolved with minimal bit errors using a standard differential receiver. The ZS results fall somewhere in the middle and are interesting for two primary reasons. First, the ZS configurations clearly outperform the traditional single-ended signaling, indicating that controlling the balance of 0's and 1's within a wide, single-ended, high-speed bus does improve the eye opening at the receiver in a realistic full link environment.

The second noteworthy observation requires a bit more discussion. For both "worst case" and "typical" operation, the ZS eye openings are almost exactly one-half that of the DIFF results. Because the ideal differential eye opening is theoretically 2x single-ended, this observed relationship between ZS and DIFF may seem believable. However, it is not immediately clear why the differential simulation results show a dependence on the pair-pair pattern alignment ("worst case" vs. "typical"). Additionally, traditional differential signaling should be more immune to (and create less) common-mode noise such as SSN. Therefore, the expectation is that the differential vertical eye openings should be *more* than twice the ZS results. To explore this issue further, the power supply noise measurements (probe located at node 'c', as shown in Figure 4) are plotted in Figure 9 for each of the three architectures.

These measurements indicate relatively high ripple observed for the differential, "worst case" configuration. Theoretically, truly differential signals are inherently zero sum (or zero current), thus the power supply ripple should be somewhat lower despite the pair-pair pattern relationships. However, the use of two independent push-pull style SST buffers with minimal on-die capacitance, each terminated to a common-mode voltage, results in the quasi-differential buffer scheme. These deviations from traditional differential signaling may have contributed to localized rail collapse and higher ripple when operating with "worst case" patterns. If further study is considered, it may become necessary to emulate a more traditional differential network where the resulting eye opening may be more optimistic.

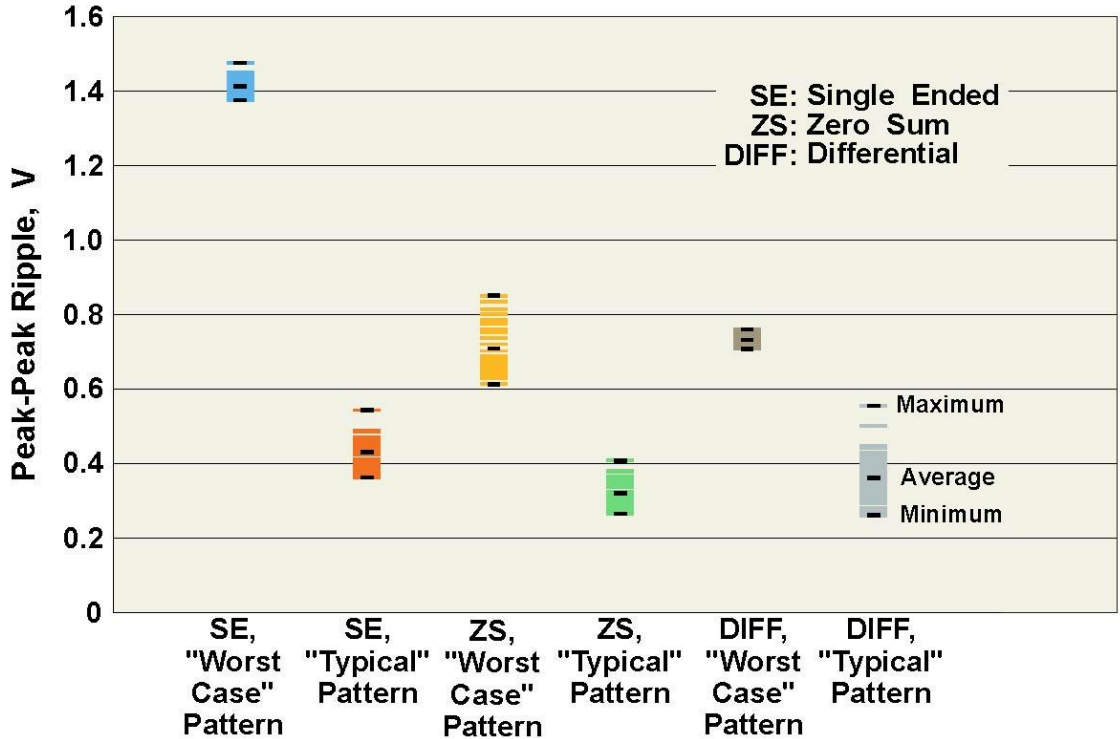

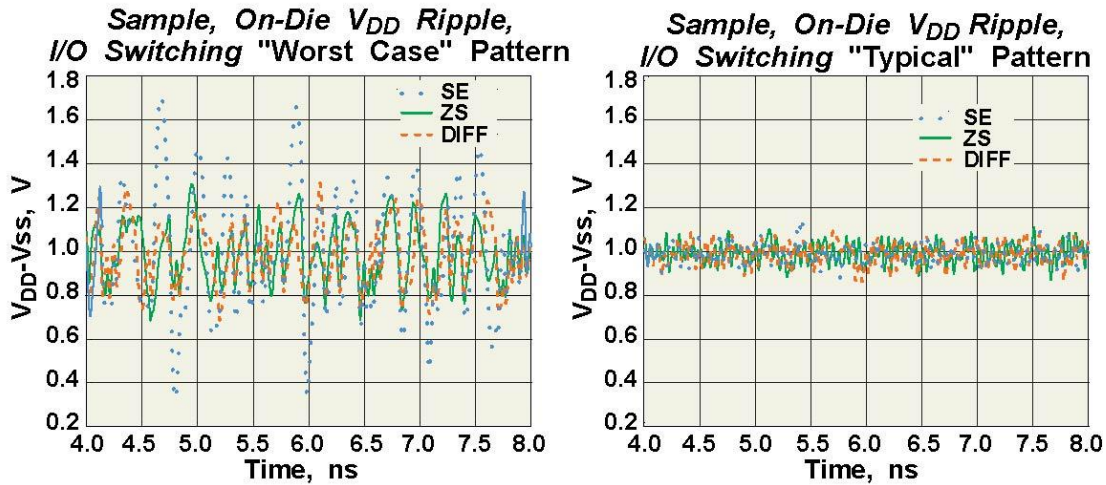

Figure 9: $V_{DD}$ On-Die Power Supply Noise for Three Buffer Configurations and Two Switching Patterns In a 16 Gbps Simulation. (41296)

The simulated VDD voltage ripple at the die was sampled with respect to the local, non-ideal VSS reference (node 'd' in Figure 4). Even though VDD is only 1 VDC, it is possible to observe peak-peak AC ripple greater than 1 V; which is exactly the situation for the SE "worst case" condition. The voltage seen by the buffer cell is 1 VDC +/- 700 mV. Again, it is no surprise the SE "worst case" vertical eye opening was so small considering the input voltage available for each buffer.

## Simulation Variation: Disparity and Bus Size

Limiting the available code words to purely balanced (zero sum) codes may be too constraining in a real application. The concept of disparity allows some finite amount of imbalance between the number of 0's and 1's across the bus, as discussed in the principles of operation section. This makes more code words available from the possible $2^N$ complete list at a cost of incrementally increasing SSN. When the code space is widened in this manner, fewer wires are required to achieve 32 logical bits. For example, with a disparity of +/-2, only 34 wires are required for 32 logical bits versus the 36 bits required in a purely balanced encoding. The zero sum I/O slice and bit patterns were modified to include varying levels of disparity to visualize the tradeoff between increasing the code space with unbalanced code words and the impact to the eye opening caused by the increased SSN. As shown in the left panel of Figure 10, increasing the disparity does gradually close the eye; small levels may be tolerable if the code word limitation can be relaxed. It is noted here that opening the code space to allow a disparity of +/-16 requires only 32 wires; effectively resulting in un-encoded words. However, we chose to maintain the 34 wire example with a +/- 16 disparity for the sake of consistency.

Similarly, alternative bus sizes were considered as another variation on the baseline application. Rather than a single, zero sum bus consisting of 36 wires to get 32 logical bits, a 2x20 and 4x12 configuration were also evaluated; each again achieving 32 bits. There appears to be minimal impact to the eye opening when considering these two additional architectures.

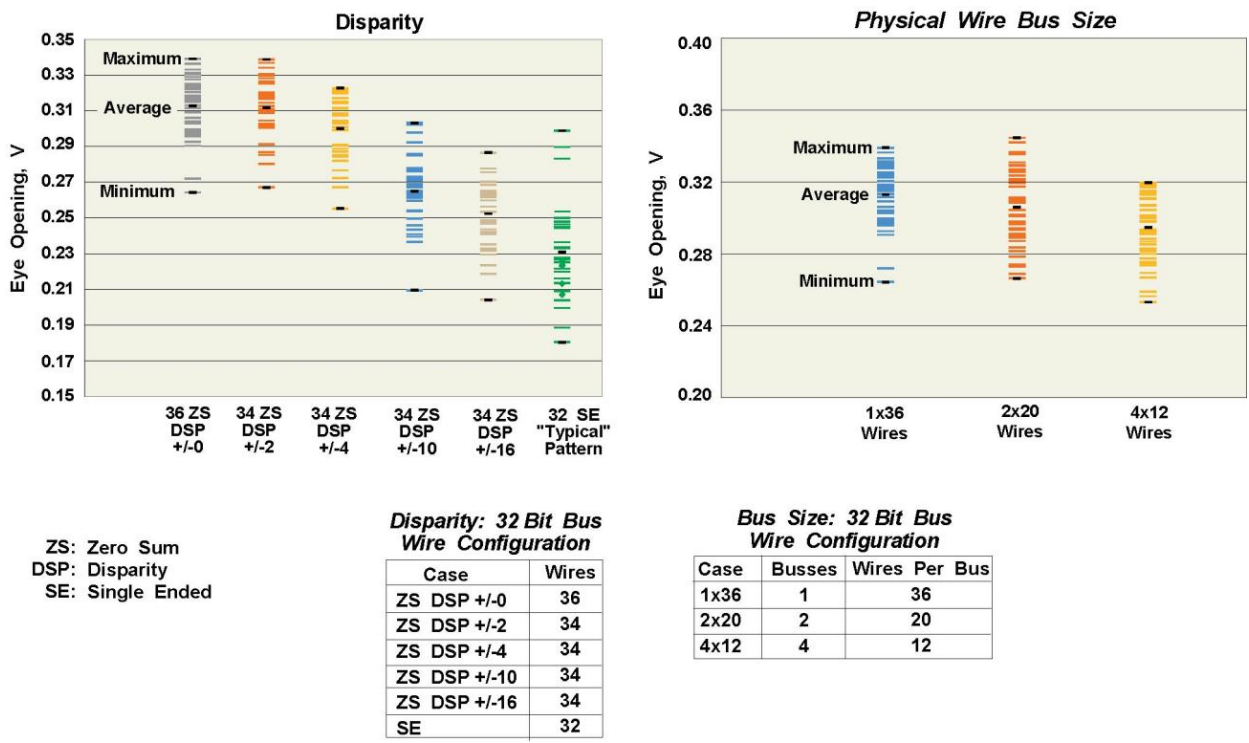

Figure 10: Disparity and Bus Size Impact on Eye Opening. (41293)

## Simulation Variation: Data Rate

Because SSN is a frequency-dependent power integrity issue, it was speculated that the performance of the single-ended signals would begin to degrade at approximately 1 Gbps due primarily to the inductance in the via pinfield and on-die PDN.  To verify this assertion, the transmit data rate was stepped from 233 Mbps to 16 Gbps.  The results from this sweep are shown in Figure 11.  For the single-ended, "worst case" configuration, the eye is approximately 50% closed between 1 and 2 Gbps, whereas the zero sum eyes remain as open as that of the differential eyes throughout the complete frequency sweep.  (Note that the differential eye measurements were halved for comparative plotting purposes.)  Even for applications with data rates as low as 1 Gbps, the zero sum coding approach is more immune to the negative impact of SSN as compared to traditional single-ended signals.

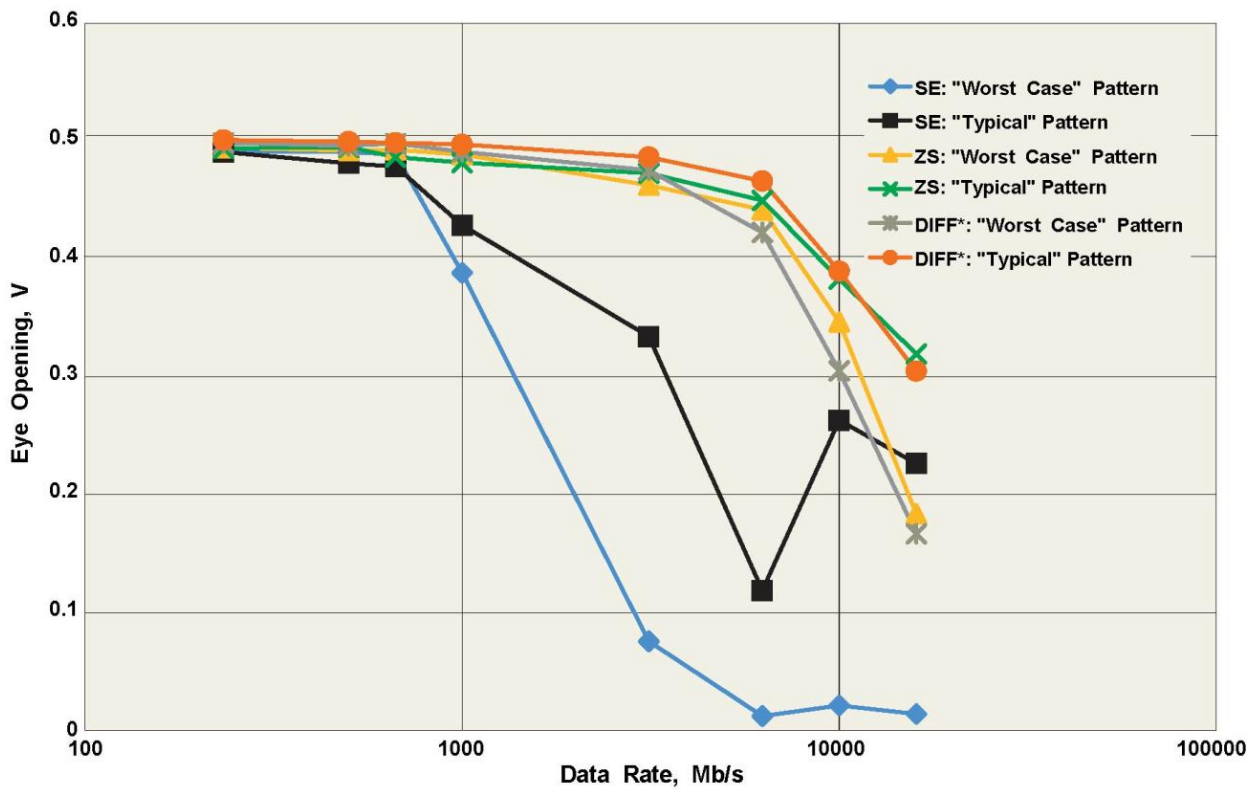

Figure 11:  Impact of Data Rate on Vertical Eye Opening.  (41290)

## Zero Sum Signaling Conclusions

The simulation results presented in this report suggest that SSN immunity is realizable in a realistic system environment when using a zero sum signaling configuration.  As expected, equally balancing the number of ones and zeroes across a wide data bus results in current cancellation which subsequently causes reduced voltage rail collapse and improved vertical eye opening at the receivers.  This performance improvement is also typically achieved using differential signals, but at the penalty of 2x the packaging density over a single-ended approach. Zero sum signaling appears to be able to achieve good signaling performance with a significantly smaller pin-count and routing penalty.

It is worth restating that the DIFF simulations performed were quasi-differential in that a true current-steering differential transmitter was not used. Such a differential buffer could have produced more optimistic vertical eye openings.  Similarly, a significantly higher level of on-die capacitance for local charge storage could likely improve the results for all signaling styles.

While the simulation results appear promising for reducing SSN in a high-speed, parallel bus HPC application using zero sum signaling, there are several remaining issues to be considered. For example, the design and implementation of an efficient encoding/decoding scheme is critical for advancing the concept of zero sum signaling into a physical reality.  We plan to report on this topic in future publications.  Compatibility with traditional system architectures (often requiring DC-free and/or ECC coding) may also be important.

Additionally, we note the processor-processor notional application diagram shown in Figure 3 is conceptual.  There will certainly be architectural complications to consider.  For instance, in the case of a processor-to-memory architecture, integrating a custom coded data bus with commercial memory (e.g. DDR2/3) will be a logistical challenge.  Also, as zero sum signaling implicitly assumes synchronous clocking of all data bits within a zero sum bus, whereas recent trends in SerDes technology are quite the opposite (separate SerDes channels often sync to separate PLLs and hence are not synchronous), compatibility with future high-speed I/O solutions would need to be addressed.